\documentclass[twocolumn,aps,prd,showpacs,nofootinbib,eqsecnum,amsmath,amsfonts,amssymb]{revtex4}
\usepackage[dvips]{graphicx}

\newcommand{\Long}{\mathbb{L}}

\def\pl#1#2#3{ Phys. Lett. {\bf #1}, #2 (#3)}
\def\prl#1#2#3{ Phys. Rev. Lett. {\bf #1}, #2 (#3)}
\def\prd#1#2#3{ Phys. Rev. D {\bf #1}, #2 (#3)}
\def\pr#1#2#3{ Phys. Rev. {\bf #1}, #2 (#3)}

\def\cqg#1#2#3{ Class. Quantum Grav. {\bf #1}, #2 (#3)}
\def\pu#1#2#3{ Physics-Uspekhi {\bf #1}, #2 (#3)}
\def\ijmpd#1#2#3{ Int. J. Mod. Phys. D {\bf #1}, #2 (#3)}
\def\ptp#1#2#3{ Prog. Theor. Phys. {\bf #1}, #2 (#3)}

\begin{document}
 
\title{Conformal-thin-sandwich initial data for a\\
       single boosted or spinning black hole puncture}
 
\author{Pablo Laguna}
\affiliation{Center for Gravitational Physics and Geometry, \\
Center for Gravitational Wave Physics,\\
Department of Astronomy and Astrophysics,\\
Department of Physic,\\
Penn State University, University Park, PA 16802}
 
\begin{abstract}
Sequences of initial-data sets representing binary black holes in 
quasi-circular orbits have been used to calculate what may be interpreted
as the innermost stable circular orbit. 
These sequences have been computed with two approaches.
One method is based on the traditional conformal-transverse-traceless decomposition
and locates quasi-circular orbits from the turning points in an effective potential.  
The second method uses a conformal-thin-sandwich  
decomposition and determines quasi-circular orbits by requiring
the existence of an approximate helical Killing vector. 
Although the parameters defining the innermost stable circular orbit obtained from these
two methods differ significantly, both approaches yield approximately the 
same initial data, as the separation of the binary system increases. 
To help understanding this agreement between data sets, 
we consider the case of initial data representing a single boosted or 
spinning black hole puncture of the Bowen-York type
and show that the conformal-transverse-traceless and conformal-thin-sandwich 
methods yield identical data, both satisfying the conditions for the
existence of an approximate Killing vector.
\end{abstract}
 

\maketitle

\section{Introduction}

Binary systems involving compact objects such as black holes and neutron stars 
have been the main source of attention of numerical relativists in the past two
decades. The motivation for studying these systems has been enhanced by the urgency
of producing results that could help observational efforts like LIGO,
GEO600, TAMA300 and VIRGO \cite{detectors}. Unfortunately the problem of producing
complete and sufficiently accurate numerical simulations of orbits and mergers
of compact binaries is still not feasible. This impediment does not imply that the results produced so far
are devoid of any astrophysical relevance. Simulations of
a few orbits of neutron star binaries have been obtained \cite{ns_orbits,pedro}.
Grazing collisions and wave forms produced from the final plunge of a binary black hole system have been
also computed \cite{bbh_waves}. And of direct relevance to the work presented in this paper is
the study of the behavior of binary orbits using sequences of initial data configurations.
 
Astrophysically realistic initial data sets representing black hole binaries are crucial.
Without them, the predictive power of gravitational wave source simulations,
with relevance to data analysis efforts, is
seriously compromised.  Constructing initial data sets involves a 
series of choices such as conformal transformations, tensor decompositions, 
topologies, boundary conditions, freely specifiable data to name a few. 
These choices have a profound
effect on the physical properties of the data.

Two general approaches for constructing initial data are currently under 
consideration. The oldest approach has its roots in the work by Lichnerowicz and York \cite{york79}.
This approach is based on conformal transformations and transverse-traceless (CTT) decompositions.
The second and most recent approach, currently receiving considerable attention,
attempts to establish a more direct ``space-time" connection.
It borrows some of the ideas from the CTT method, namely conformal transformations
and traceless decompositions, but in addition provides a natural framework for setting up
quasi-equilibrium configurations of binary systems. 
Under this approach, initial data can be 
viewed as derived from a ``thick slice or thin sandwich" section 
of the space-time. The method was pioneered by
Mathews and Wilson \cite{wilson}; recently used by Grandcl\'ement, Gourgoulhon and 
Bonazzola \cite{ggb1}; and formally characterized by York \cite{york99}.
Because the approach also involves conformal transformations, it is known as 
the conformal thin-sandwich (CTS) method.

Initial data sets representing binaries in quasi-circular orbits will be needed as a
starting point
of any evolution simulation aimed at producing gravitational waveforms of astrophysical
black hole binaries. Cook \cite{cook} introduced the first method to locate what could be
identified as black hole binaries in quasi-circular orbits. The method, called
the effective potential method, is based on the potential 
\begin{equation}
\label{epot}
E_b = E - 2\,M\,,
\end{equation} where $E_b$ represents the binding energy of the system,
$E$ the total energy and $M$ the irreducible mass of each individual black hole. 
The total energy $E$ is a well defined quantity, namely the total ADM mass
\cite{ADM}. However, 
defining individual black hole masses
for interacting black holes is not a completely well defined concept \cite{isolated}.
Nonetheless, a reasonable approximation is to assume that the irreducible mass is given
by the proper area of the apparent horizon. Quasi-circular orbits are then found from the
condition
\begin{equation}
\frac{\partial E_b}{\partial l}\Big|_{M,J} = 0\,,
\end{equation}
where $l$ is the proper separation of the horizons and $J$ the total angular momentum of the
binary. Similarly, the angular velocity $\Omega$ is found from
\begin{equation}
\Omega = \frac{\partial E_b}{\partial J}\Big|_{M,l}\,.
\end{equation}
Given this approach to identifying quasi-circular orbits, the innermost stable circular orbit (ISCO)
is found by looking at the end point of a sequence of quasi-circular orbits.

Baumgarte \cite{baumgarte} and later Baker \cite{baker} recomputed, using Cook's effective
potential method, quasi-circular orbit sequences.   
One of the motivations for this work was to investigate whether Cook's results were
sensitive to how the black hole singularities are handled. 
Instead of using a topology requiring conformal-imaging as used by Cook \cite{cook},
Baumgarte and Baker handled the black hole singularities via the
puncture method \cite{puncture}. Both studies 
produced results in close agreement with those by Cook. 
In the remainder of this paper, quasi-circular orbit data sets computed with the CTT method
will be referred as Cook-Baumgarte-Baker (CBB) data sets. 

As mentioned before, the CTS approach has also been used by
Gourgoulhon, Grandcl\'ement and Bonazzola (GGB) 
\cite{ggb1,ggb2} 
to construct quasi-circular orbits of black hole binary systems. 
The essence of this work is to identify quasi-circular orbits
based on the assumption that the space-time possesses an approximate 
helical Killing vector. 
One important aspect of the GGB sequences is that they
yield an ISCO identification in better agreement with 
post-Newtonian results \cite{postNewt} when compared with CBB sequences.
Specifically, the normalized binding energy $e = E/2\,M - 1$ at the ISCO from
post-Newtonian estimates is $e \approx -1.67\%$. The GGB study yields 
$e \approx -1.73\%$. On the other hand, the CBB estimate is significantly more 
strongly bound, 
$e \approx -2.3\%$. These differences show more dramaticaly when comparing the ISCO orbital period.
Post-Newtonian calculations yield $T/2\,M \approx 71.2$ and in the GGB work 
$T/2\,M \approx 62.8$. In contrast, the CBB data estimate $T/2\,M \approx 37$.

Reasons for these discrepancies are not completely 
understood. One suggestion was that these differences
were due to the criteria for identifying quasi-circular orbits. 
Instead of the method used by CBB described above,
GGB determined the co-rotating frame defining quasi-circular orbits 
from the requirement that the ADM and Komar \cite{Komar}
masses are equal. The motivation for this choice was that the Komar and ADM masses only agree
for stationary space-times. Therefore, the data computed this way have a better chance of
representing a quasi-stationary situation. Skoge and Baumgarte \cite{skoge} designed a
simple example involving thin-spherical shells of collision-less particles in circular 
orbits which shows that the effective potential method and the equal Komar and ADM masses 
method for identifying circular orbits are equivalent.
More recently, Tichy and Br\"ugmann \cite{tichy} computed sequences of 
CTS binary black hole punctures that yield a similar conclusion.
These studies provide strong support to the view that the differences between CBB and GGB
data sets are not due to the method for identifying quasi-circular orbits but somewhere else.

Another possible explanation for the ISCO discrepancy that 
has been mentioned is once again the approach to representing the 
black hole singularities. The GGB initial data sequences are based on a topology
requiring conformal-imaging \cite{cook}. 
The work in Refs.~\cite{BLT,tichy} was mostly aimed at testing this
by extending the GGB
results to the case of black holes represented by punctures. Unfortunately, the data obtained
in this work only approximately satisfied one of the conditions
used by GGB to define quasi-stationarity. Recently, Hannam and collaborators \cite{hannam}
pointed out the serious difficulties in  
constructing CTS binary data in quasi-equilibrium 
based on punctures instead of conformal-imaging.  
   
There is however on important aspect of the
CBB and GGB sequences that cannot be ignored. 
In spite of their differences for determining the ISCO, the CBB and GGB sequences show 
a close agreement when the separation of the binary systems increases.
Therefore, at large separations, it should be possible to translate CBB into GGB data and vice-versa,
independently of the procedure used to represent the black holes.    
The work in this paper is motivated by this observation.
Since for large binary separations the black holes
are approximately isolated, we focus 
attention on single boosted or spinning black hole punctures. 
We demonstrate that it is possible to translate 
the CTT data into a CTS form while preserving the conditions used by GGB, namely
the existence of an approximate Killing vector, thus circumventing in these cases 
the problems pointed out in Ref.~\cite{hannam}.

The paper is organized as follows.
In Sec.~\ref{sec:decomp}, the conformal decomposition of the Einstein constraints is presented up to
the point where the CTT and CTS methods agree. Brief reviews of the CTT and CTS methods are
presented in Secs~\ref{sec:ctt} and \ref{sec:cts}, respectively. The conditions necessary for
translating between CTT and CTS data are derived in Sec.~\ref{sec:recon}. The main results
regarding boosted or spinning punctures are presented in Sec.~\ref{sec:punctures}.
Conclusions are given in Sec.~\ref{sec:conclusions}.

\section{Decomposition of the Constraints}
\label{sec:decomp}

Under the standard 3+1 decomposition of the Einstein equations, the 4-dimensional
line element is written as
\begin{equation}
\label{eq:ds}
ds^2 = -\alpha^2\,dt^2 + \tilde g_{ij}\,(dx^i+\beta^i\,dt)\,(dx^j+\beta^j\,dt)\,,
\end{equation}
with $\tilde g_{ij}$ the intrinsic 3-metric to $\Sigma_t$ ($t$ = constant hypersurfaces), $\alpha$ the
lapse function and $\beta^i$ the shift vector.\footnote{Latin 
indices denote 3-dimensional spatial indices and Greek indices 4-dimensional space-time indices.}
The extrinsic curvature of $\Sigma_t$ is defined by
\begin{equation}
\label{eq:Kij}
\tilde K_{ij} \equiv -\frac{1}{2}\,{\cal L}_N\,\tilde g_{ij}\,,
\end{equation}
with ${\cal L}_N$ the Lie derivative along the unit time-like normal 
$N^\mu$ to $\Sigma_t$.
In terms of these variables, the vacuum Einstein equations consist \cite{ADM} of a set of evolution
equations:
\begin{eqnarray}
\partial_o\, \tilde g_{ij} 
&=& - 2\, \alpha\, \tilde K_{ij}
\label{gdot} \\
\partial_o\, \tilde K_{ij}
&=&  - \widetilde\nabla_i
\widetilde\nabla_j \alpha +   \alpha\,\tilde R_{ij} \nonumber\\
& +& \alpha\,(\tilde K\, \tilde K_{ij} - 2\, \tilde K_{ik}\, \tilde K^k\,_j) \,,\label{Kdot}
\end{eqnarray}
and a set of constraint equations:
\begin{eqnarray}
\tilde R + \tilde K^2 - \tilde K_{ij} \tilde K^{ij} &=& 0  \label{hamcons}\\
\widetilde\nabla_j \tilde K^{ij} - \widetilde\nabla^i \tilde K & =& 0 \, .\label{momcons}
\end{eqnarray}
Above, $\widetilde\nabla_i$ and $\tilde R_{ij}$ denote respectively covariant differentiation
and Ricci tensor associated with $\tilde g_{ij}$. Also 
$\tilde K\equiv\tilde g^{ij}\tilde K_{ij}$, 
$\tilde R\equiv\tilde g^{ij}\tilde R_{ij}$ and $\partial_o \equiv \partial_t - {\cal L}_{\beta}$.

Constructing initial data involves specifying ($\tilde g_{ij},\, \tilde K_{ij}$)
on a given $\Sigma_t$ subject to the four constraint equations (\ref{hamcons}) and
(\ref{momcons}). In other words, out of the twelve pieces 
in ($\tilde g_{ij},\, \tilde K_{ij}$), only eight are freely specifiable. The remaining four
are fixed by these constraints. Thus, 
setting up the initial data problem in general relativity becomes
the task of singling out in ($\tilde g_{ij},\, \tilde K_{ij}$) those quantities that will
be fixed by the constraints. This is accomplished via conformal transformations
and tensorial decompositions. The CTT and CTS approaches are examples of this
methodology. 

Both the CTT and CTS methods involve applying a conformal transformation of the
3-metric and a decomposition of the extrinsic curvature of the form:
\begin{eqnarray}
\label{eq:gconf}
\tilde g_{ij} &=& \phi^4\,g_{ij} \\
\label{eq:Kdecomp}
\tilde K_{ij} &=& \tilde A_{ij}+\frac{1}{3}\,\tilde g_{ij}\, K\,,
\end{eqnarray}
such that $\tilde A^i\,_i = 0$. Notice that the choice
$\tilde K = K$ has been made.
In terms of these new variables, the constraint equations 
(\ref{hamcons}) and (\ref{momcons}) become:
\begin{eqnarray}
\phi^{-4}\, R &-& 8\,\phi^{-5}\,\nabla^2\phi 
       + \frac{2}{3} K^2 - \tilde A_{ij} \tilde A^{ij}  = 0 \label{eq:hc}\\
\nabla_j \tilde A^{ij} &+& 10\,\tilde A^{ij}\,\nabla_j\ln{\phi} 
- \frac{2}{3}\,\phi^{-4}\,\nabla^i\, K = 0\,,\label{eq:mc}
\end{eqnarray}
where $\nabla_i$ and $R$ denote respectively covariant differentiation
and Ricci scalar associated with the conformal 3-metric $g_{ij}$.
In deriving Eq.~(\ref{eq:mc}), we have used the following identity:
\begin{equation}
\label{eq:rule}
\widetilde\nabla_j\,\tilde T^{ij} = \phi^n\,[\nabla_j T^{ij} 
+ (10+n)\,T^{ij} \nabla_j\ln{\phi}]\,,
\end{equation}
where $\tilde T^{ij} = \phi^n\, T^{ij}$, with $n$ an exponent.
Similarly, the evolution equations (\ref{gdot}) and
(\ref{Kdot}) can be rewritten as:
\begin{eqnarray}
\partial_t \phi &=& \beta^i\,\nabla_i\,\phi
+ \frac{1}{6}\,\phi\,\left(\nabla_i\beta^i -\alpha\,K \right)\label{phidot}\\ 
\partial_t g_{ij} &=& 
(\Long\,\beta)_{ij}- 2\,\alpha\,\phi^{-4}\,\tilde A_{ij} \label{etadot}\\
\partial_o \, K &=& -\widetilde\nabla^2\,\alpha
+ \alpha\,\left(\tilde A^{ij}\,\tilde A_{ij} + \frac{1}{3}K^2\right)\label{trkdot}\\
\partial_o\,\tilde A_{ij} &=& [-\widetilde\nabla_i\widetilde\nabla_j\alpha 
+\alpha\,\tilde R_{ij}]^{TF} \nonumber\\
&-& \alpha\left(2\,\tilde A_{ik}\,\tilde A^k\,_j
+ \frac{1}{3}\,K\,\tilde A_{ij}\right)\label{adot}
\end{eqnarray}
where the superscript $TF$ denotes the trace-free part of the tensor within
square brackets and 
\begin{equation}
\label{eq:long}
(\Long \beta)_{ij} \equiv   2\,\nabla_{(i}\,\beta_{j)} - \frac{2}{3}\,g_{ij}\,\nabla_k\,\beta^k\,.
\end{equation}

Since we are focusing our attention on the methodologies followed by CBB and GGB, we
introduce at this point the assumptions that both share, namely conformal flatness 
$g_{ij}=\eta_{ij}$ and
vanishing of $K$. These two conditions exhaust six of the eight freely specifiable 
quantities in ($\tilde g_{ij},\, \tilde K_{ij}$). Under these assumptions,  Eqs.~(\ref{eq:hc}) and
(\ref{eq:mc}) reduce to:
\begin{eqnarray}
&&\nabla^2\,\phi   +\frac{1}{8}\,\phi^5\tilde A_{ij} \tilde A^{ij}=0 \label{eq:hc2}\\
&&\nabla_j \tilde A^{ij} + 10\,\tilde A^{ij}\,\nabla_j \ln{\phi} = 0\,.\label{eq:mc2}
\end{eqnarray}

In the remainder of the paper, 
we will continue referring to the approach followed by CBB as CTT and by GGB as
CTS, but one should keep in mind that CTT and CTS are more general. They 
do not necessarily require conformal flatness nor $K=0$.

\subsection{Conformal-Transverse-Traceless Decomposition}
\label{sec:ctt}

The CTT method was pioneered by York and collaborators \cite{york79}. Until recently, 
CTT has been the preferred method to construct initial data in numerical relativity. 
Given Eqs.~(\ref{eq:hc2}) and (\ref{eq:mc2}), one introduces the following
conformal transformation and transverse-longitudinal decomposition of the
extrinsic curvature:
\begin{eqnarray}
\label{Actt}
\tilde A^{ij} &=& \hat\phi^{-10}\,\hat A^{ij} \nonumber \\
              &=& \hat\phi^{-10}\,[\Theta^{ij} + (\Long W)^{ij}]\,,
\end{eqnarray}
such that
$\nabla_j\,\Theta^{ij} = 0$ and $\Theta^i\,_i = 0$.
We label CTT quantities with hats.

Here $\Theta^{ij}$ represents the transverse part of $\hat A^{ij}$. Since $\Theta^{ij}$ is
trace and divergence free, it contains the two remaining 
freely specifiable pieces of the initial data ($\tilde g_{ij},\, \tilde K_{ij}$). The longitudinal part of
$\hat A^{ij}$ is given by $(\Long W)^{ij}$. 
It is this part of $\hat A^{ij}$ that is fixed by the
momentum constraint (\ref{eq:mc2}). Furthermore, it is not difficult to show
that $\Theta^{ij}$ and $(\Long W)^{ij}$ satisfy the following orthogonality condition:
\begin{equation}
\label{ortho}
\int_{\Sigma_t} \Theta^{ij}\,(\Long W)_{ij} \sqrt{\eta}\,d^3x = 0\,.
\end{equation}

For simplicity CBB assumed that $\Theta^{ij} = 0$. Therefore, substitution of (\ref{Actt})
into (\ref{eq:hc2}) and (\ref{eq:mc2}) yields:
\begin{eqnarray}
&&\nabla^2\,\hat\phi  +\frac{1}{8}\,\hat\phi^{-7} 
(\Long W)^2 = 0\label{ctt:hc}\\
&&\nabla_j (\Long W)^{ij} = 0\,,\label{ctt:mc}
\end{eqnarray}
where we have introduced the notation 
$(\Long W)^2 \equiv (\Long W)^{ij}\,(\Long W)_{ij}$.
In summary, the initial data problem under the CTT approach reduces to solving 
first Eq.~(\ref{ctt:mc}) for $W^i$ 
and then using this solution to solve for $\hat\phi$ from Eq.~(\ref{ctt:hc}).
 
Bowen and York \cite{BY} found solutions to the momentum constraint (\ref{ctt:mc}) 
representing single black holes with linear momentum $P^i$ or 
angular momentum $J^i$. Specifically,
\begin{eqnarray}
\label{wlin}
W^i &=& -\frac{1}{4\,r} (7\,P^i + n^i\,n_j\,P^j) \\
\label{wspin}
W^i &=& -\frac{1}{r^2}\,\epsilon^{ijk}\,J_j\,n_k\,,
\end{eqnarray}
with $n^i$ the unit normal of constant $r$ spheres in flat space;
namely, $n^i = (x,\, y,\, z)/r = (\partial/\partial\,r)^i$ and
$r = ||x^i||$. 
In terms of the solutions (\ref{wlin}) and (\ref{wspin}),
the extrinsic curvature $\hat A^{ij}$ takes the form:
\begin{eqnarray}
\label{alin}
\hat A^{ij} &=& \frac{3}{2\,r^2}\left[ 2\,P^{(i}\,n^{j)} - (\eta^{ij} - n^i\,n^j)\,n_k\,P^k\right] \\ 
\label{aspin}
\hat A^{ij} &=& \frac{6}{r^3}\,n^{(i}\,\epsilon^{j)kl}\,J_k\,n_l \,. 
\end{eqnarray}

Given the solution with linear momentum (\ref{alin}), solutions 
representing non-spinning binary black holes are obtained via the following superposition:
\begin{equation}
\label{eq:superposition}
\hat A^{ij} = \hat A^{ij}(C_1^k,\,P^k_1) + \hat A^{ij}(C_2^k,\,P^k_2)\,,
\end{equation}
where $C^i_A$ with $A=1,\,2$ denotes the coordinate location of each of the
black holes. The explicit expression for $\hat A^{ij}(C_A^i,\,P^i_A)$ 
is given by (\ref{alin}) with $r_A = ||x^i-C^i_A||$ and
$n^i_A = (x^i-C^i_A)/r_A$. It is not difficult to show that, with 
$\hat A^{ij}$ given by (\ref{eq:superposition}), the total
ADM linear and angular momentum (about the origin of the coordinate system) are
$P^i = P^i_1+P^i_2$ and $J_i = \epsilon_{ijk}\,(C^j_1\,P^k_1 + C^j_2\,P^k_2)$,
respectively. 

CBB specialized to the case of equal 
mass black holes, in the center-of-momentum frame of the binary system
and in quasi-circular orbit. Thus, $P^i \equiv P^i_1 = -P^i_2$ and 
$P_i\,C^i \equiv P_i\,(C^i_1-C^i_2)= 0$.
In addition, asymptotic flatness
requires that in solving the Hamiltonian constraint (\ref{ctt:hc}) one imposes the following 
Robin boundary condition as $r\rightarrow \infty$:
\begin{equation}
n^i\,\nabla_i\hat\phi = \frac{(1-\hat\phi)}{r}\,.
\end{equation}
The differences between
the procedures followed by Cook \cite{cook} and by Baumgarte \cite{baumgarte} and Baker \cite{baker}
entered in the approach used to handle the black hole singularities.
Cook \cite{cook} used the conformal-imaging approach.
Baumgarte \cite{baumgarte} and Baker \cite{baker}, on the other hand, applied the puncture
method developed by Brandt and Br\"ugmann \cite{puncture}.
Constructing sequences of initial data sets reduces then to a 3-dimensional parameter space consisting of
$C \equiv ||C^i||$, $P \equiv ||P^i||$ and a third parameter. For the conformal-imaging approach
method the third parameter is related to the
radius of the throat of the black holes and for the puncture method to their ``bare" mass.

\subsection{Conformal-Thin-Sandwich Decomposition}
\label{sec:cts}

The fundamental difference between the CTT method used by GGB and the CTS method used by CBB 
lies in the decomposition of the extrinsic curvature. Instead of using (\ref{Actt}), one introduces
\begin{eqnarray}
\label{eq:Acts}
\tilde A^{ij} &=& \bar\phi^{-4}\,\bar A^{ij} \nonumber \\
              &=& \bar\phi^{-4}\,\frac{1}{2\,\alpha}\,[- u^{ij} + (\Long \beta)^{ij}]\,,
\end{eqnarray}
where bars will be used to denote CTS quantities.

The first aspect to notice is that, strictly speaking, $(\Long \beta)^{ij}/2\,\alpha$
and $u^{ij}/2\,\alpha$ cannot be viewed respectively as the 
longitudinal and transverse parts of $\bar A^{ij}$. That is, they do not satisfy the
orthogonality condition (\ref{ortho}). In Ref.~\cite{PY}, the measure used in (\ref{ortho})
is modified to show that it is possible to arrive to a orthogonal decomposition of $\bar A^{ij}$
of the form
\begin{equation}
\bar A^{ij} = \Theta^{ij} + \frac{1}{2\,\alpha}\,(\Long B)^{ij}\,,
\end{equation}
with $\nabla_j\Theta^{ij} = \Theta^i\,_i = 0$. The orthogonality condition in this case reads
\begin{eqnarray}
\label{ortho2}
\int_{\Sigma_t} \Theta^{ij}\,\frac{1}{2\,\alpha}\,(\Long B)_{ij}\,2\,\alpha\, \sqrt{\eta}\,d^3x &=&  \nonumber \\
\int_{\Sigma_t} \Theta^{ij}\,(\Long B)_{ij}\,\sqrt{\eta}\,d^3x &=& 0\,.
\end{eqnarray}

The next step in the CTS approach is to realize that the freely specifiable nature
of $u_{ij}$ can be exploited to find quasi-equilibrium configurations. 
From the evolution equation (\ref{etadot}) one has that
\begin{equation}
\label{eq:ctscond}
\partial_t \bar g_{ij} = (\Long \beta)_{ij} - 2\, \alpha\, \bar A_{ij} = u_{ij}\,.
\end{equation}
For quasi-equilibrium configurations
defined by an approximate helical Killing vector, 
it is then natural to choose $u_{ij} = 0$.

With $u_{ij} = 0$, the Hamiltonian and momentum constraint equations (\ref{eq:hc2}) and (\ref{eq:mc2}) 
read
\begin{eqnarray}
&&\nabla^2\,\bar\phi  +\frac{\bar\phi^5}{32\,\alpha^2} (\Long \beta)^2 = 0\label{cts:hc}\\
&&\nabla_j (\Long \beta)^{ij} - (\Long\beta)^{ij}\,\nabla_j\ln{(\alpha\,\bar\phi^{-6})} = 0\,.\label{cts:mc}
\end{eqnarray}
Because of the particular form used to decompose $\bar A_{ij}$ in (\ref{eq:Acts}), 
a recipe for fixing $\alpha$ is also needed. Since one of the assumptions used 
is $K = 0$, it is natural to require for 
$\Sigma_t$ to be a maximal slice, namely $\partial_t\, K = 0$.
Substitution of $K = \partial_t\,K = 0$ into
the evolution equation (\ref{trkdot}) yields 
\begin{equation}
\nabla^2\,\alpha + 2\,\bar\phi^{-1}\,\nabla_i\bar\phi\,\nabla^i\alpha = \frac{\bar\phi^4}{4\,\alpha} 
(\Long \beta)^2\,, \label{eq:lapse}
\end{equation}
from which one can solve for $\alpha$.
It is convenient to rewrite (\ref{eq:lapse}) with the help
of Eq.~(\ref{cts:hc}) as follows: 
\begin{equation}
\nabla^2(\alpha\bar\phi) -  
\frac{7}{32}\,\frac{\bar\phi^6}{(\alpha\,\bar\phi)}\, (\Long \beta)^2 = 0\label{eq:lapse2}\,.
\end{equation} 

Notice from (\ref{phidot}) and (\ref{adot})that choosing
$u_{ij} = 0$, $\partial_t K = 0$ and $K = 0$ does not necessarily 
imply that $\partial_t\bar\phi = 0$ nor
$\partial_t \bar A_{ij} = 0$, which would be necessary in order to have an exact
helical Killing vector.

The boundary conditions needed to solve Eqs.~(\ref{cts:hc}), (\ref{cts:mc}) and (\ref{eq:lapse}) 
are that as $r\rightarrow \infty$,
\begin{eqnarray}
n^i\,\nabla_i\alpha &=& \frac{(1-\alpha)}{r} \\
n^i\,\nabla_i\bar\phi &=& \frac{(1-\bar\phi)}{r} \\
\beta^i &\rightarrow& \Omega\,\left(\frac{\partial}{\partial\varphi_o}\right)^i\,,
\end{eqnarray} 
with $\Omega$ the orbital angular velocity of the binary and
$(\partial/\partial\varphi_o)^i$ the Killing vector of the flat metric for rotation 
in the $\varphi_o$ direction.
In the GGB set up, the black holes are represented by two sheets of the Misner-Lindquist manifold (see
\cite{ggb1} for details). That is, there are two throats $S_1$ and $S_2$ of radii $a_1$ and $a_2$
located at coordinates $C^i_1$ and $C^i_2$ connecting the two sheets. Boundary conditions at these throats
are derived from isometries between the two sheets. These conditions are that at the surface
${\cal S}$ of each of the throats:
\begin{eqnarray}
\label{lapse_bc}
\alpha|_{\cal S} &=& 0\\
\label{shift_bc1}
\beta\,_\parallel|_{\cal S} &=& 0\\
\label{shift_bc2}
n^j\,\nabla_j\beta^i_\perp|_{\cal S} &=& 0\\
\label{phi_bc}
\left(n^i\,\nabla_i\bar\phi + \frac{\bar\phi}{2\,r}\right) \Big|_{\cal S} &=& 0 \,.
\end{eqnarray} 
Here $r$ denotes the radial coordinate distance from the center of the corresponding throat
and $n^i$ its unit normal. 
Also, $\beta^i_\perp = h^i\,_j\,\beta^j$ with $h_{ij} = \eta_{ij}-n_i\,n_j$ the induced 
metric in ${\cal S}$ and $\beta_\parallel = n_i\,\beta^i$.
Condition (\ref{lapse_bc}) on the lapse arises from imposing anti-symmetry with respect 
to the isometry. Because $\bar A^{ij} = (\Long \beta)^{ij}/2\,\alpha$, the condition
$\alpha|_{\cal S} = 0$ forces one to impose $(\Long \beta)^{ij}|_{\cal S} = 0$
in order to have regularity of the extrinsic curvature. 
The problem encountered by GGB was that the conditions (\ref{shift_bc1}) and (\ref{shift_bc2})
are not sufficient to guarantee that $(\Long \beta)^{ij}|_{\cal S} = 0$. 
CGG designed a procedure to achieve regularity on the throats; however, Cook \cite{cook2} showed 
that a consequence of this regularization is the introduction of constraint violations.

\section{CTT-CTS Mapping}
\label{sec:recon}

The objective now is to find a prescription to translate between CTT and CTS data.
Consider initial data $(\tilde g_{ij},\, \tilde K_{ij})_{CTT}$ and 
$(\tilde g_{ij},\, \tilde K_{ij})_{CTS}$ computed by the CTT and CTS
methods, respectively.  
Without loss of generality, because of the conformal flatness and $K = 0$ assumptions, 
one can set:
\begin{eqnarray}
\label{phi_match}
\tilde g_{ij}|_{CTS} &=& \epsilon^4\,\tilde g_{ij}|_{CTT} \\
\label{A_match}
\tilde A_{ij}|_{CTS} &=&  \tilde A_{ij}|_{CTT} + \bar\phi^{-2}a_{ij}\,,
\end{eqnarray}
with $a_{ij}$ a traceless tensor and $\epsilon$ a function. 
When re-written in terms of CTT and CTS quantities, these relations take the following form:
\begin{eqnarray}
\label{phi_match2}
\epsilon &=& \frac{\bar\phi}{\hat\phi}\\
\label{A_match2}
\frac{(\Long\bar\beta)_{ij}}{2\,\bar\alpha\bar\phi^{-6}}&=& 
\epsilon^2(\Long W)_{ij} + a_{ij}\,.
\end{eqnarray} 
with bars and hats denoting CTS 
and CTT quantities, respectively. 

Since by construction, 
\begin{equation}
\nabla_j(\Long W)^{ij} = 0 \hspace{0.5cm}\hbox{and}\hspace{0.5cm}
\nabla_j\left[\frac{(\Long \bar\beta)^{ij}}{2\,\bar\alpha\bar\phi^{-6}}\right] = 0\nonumber\,,
\end{equation} 
$\epsilon$ and $a_{ij}$ must satisfy the following condition:
\begin{equation}
\label{eq:yyy}
\nabla_ja^{ij} + (\Long W)^{ij}\nabla_j\epsilon^2 = 0\,.
\end{equation}
Clearly, if identical data are to be produced by CTT and CTS methods,
one would need to show that $\epsilon = 1$ and that the only solution
to (\ref{eq:yyy}) is $a_{ij} = 0$.

The main difficulty in relating CTT and CTS data is that one not only has
to establish a connection between their corresponding equations but also
their asymptotic conditions as well as how the singularities are handled.
Since CTT binary black hole initial data are computationally less demanding to obtain
(they only require solving Eq.~(\ref{ctt:hc})), a starting point in
checking whether CTT and CTS data are the same is to verify whether
the CTT data satisfy the CTS conditions $\partial_t \hat g_{ij} = 0$ and
$\partial_t K= 0$. 

\subsection{From CTT to CTS}
\label{sec:eqs}

We will concentrate attention on the
steps to bring CTT data into a CTS form.
That is, the initial data $(\tilde g_{ij},\, \tilde K_{ij})$ 
are given by the CTT method, namely
$\tilde g_{ij} = \phi^4 \eta_{ij}$ and 
$\tilde K_{ij} = \phi^{-2}(\Long W)_{ij}$, with $(\Long W)_{ij}$ 
given by (\ref{alin}) or (\ref{aspin}) and $\phi$ computed from
\begin{equation}
\nabla^2\,\phi = -\frac{1}{8}\,\phi^{-7} (\Long W)^2 \label{isco1}\,.
\end{equation}
To simplify notation, we have dropped the hats used to denote CTT quantities.

Satisfying the CTS condition $\partial_t K= 0$ is trivial since this condition
is essentially an equation for the lapse function. 
Following a procedure similar to that used in deriving (\ref{eq:lapse2}),
with $\tilde A_{ij} = \phi^{-2}(\Long W)_{ij}$ instead of 
$\tilde A_{ij} = \phi^4(\Long\beta)_{ij}/2\,\alpha$, one obtains
\begin{equation}
\nabla^2(\alpha\phi) =  
\frac{7}{8}\,(\alpha\,\phi)\,\phi^{-8} (\Long W)^2\label{isco2}\,.
\end{equation}

Asymptotic flatness imposes on 
Eqs.~(\ref{isco1}) and (\ref{isco2}) the following conditions
as $r\rightarrow\infty$:
\begin{eqnarray}
n^i\,\nabla_i\phi &=& \frac{(1-\phi)}{r} \\
n^i\,\nabla_i(\alpha\,\phi) &=& \frac{[1-(\alpha\,\phi)]}{r}\,, 
\end{eqnarray}
respectively. These conditions imply that 
\begin{eqnarray}
\label{phi_asymp}
\phi &=& 1+\frac{a}{r} + O(r^{-2}) \\
\label{alpha_asymp}
\alpha\phi &=& 1 - \frac{b}{r}+ O(r^{-2}) \,, 
\end{eqnarray}
which in turn yield
\begin{eqnarray}
\alpha &=& \frac{1 - a/r} 
{ 1+b/r} + O(r^{-2}) \nonumber \\
&=& 1-\frac{a+b}{2\,r} + O(r^{-2})\,.
\end{eqnarray}
Above, $a$ and $b$ are parameters related to
the ADM \cite{york79} and Komar \cite{Komar} masses, respectively.
In GGB, the condition for quasi-circular orbits
is found by equating these two masses.

Next is the condition $\partial_t g_{ij} = 0$. Once again, 
a derivation similar to what was used in obtaining (\ref{eq:ctscond}) yields
\begin{equation}
\label{eq:recon}
(\Long\beta)^{ij} = 2\,\alpha\,\phi^{-6}\, (\Long W)^{ij}\,.
\end{equation}

Solving (\ref{eq:recon}) as it stands is conceptually difficult.
There are five equations for the three components of the shift vector. 
Thus, solutions are only possible in the particular case 
when the tensor  $\alpha\,\phi^{-6}\, (\Long W)^{ij}$
is purely longitudinal.
A system of equations for $\beta^i$ that is not over-determined is obtained 
by taking the divergence of (\ref{eq:recon}):
\begin{equation}
\nabla_j (\Long \beta)^{ij} = 2\,(\Long W)^{ij}\,\nabla_j(\alpha\,\phi^{-6}) 
\label{iscoshift}\,.
\end{equation}
The potential problem of computing $\beta^i$ from Eq.~(\ref{iscoshift}) 
is that, in general, yields
\begin{equation}
\label{eq:uij}
(\Long \beta)^{ij} = 2\,\alpha\,\phi^{-6}\,(\Long W)^{ij} + u^{ij}\,,
\end{equation}
with $\nabla_ju^{ij} = 0$. A non-vanishing $u_{ij}$ would imply 
$\partial_t g_{ij} \ne 0$.
Therefore, the only possible way of reconciling CTT and CTS data is
if the solution to Eq.~(\ref{iscoshift}) yields $u^{ij}= 0$. In Ref.~\cite{BLT},
we showed that $u^{ij}\approx 0$ for binary black hole data with punctures at separations 
close to ISCO. 

Given the asymptotic behavior of $\phi$ and $\alpha$,  as $r\rightarrow \infty$,
Eq.~(\ref{eq:recon}) becomes
\begin{equation}
\label{beta_bc}
(\Long \beta)^{ij} = 2\,(\Long W)^{ij}\,.
\end{equation}
Therefore, $\beta^i \rightarrow 2\,W^i + B^i$, such that
$B^i$ is a conformal Killing vector,  
i.e. $(\Long B)^{ij} = 0$. For example, in the
case of binary data in a co-rotating frame,
\begin{equation}
\label{eq:beta_infty}
\beta^i \rightarrow 2\,W^i+\Omega\,\left(\frac{\partial}{\partial\varphi_o}\right)^i\,.
\end{equation}

In summary, 
given the vector $W^i$, the initial data 
is completely determined once
(\ref{isco1}) is solved. The solution to this equation does not require knowledge
of $\alpha$ and $\beta^i$.
The role played
by Eqs.~(\ref{isco2}) and (\ref{eq:recon}) is to provide coordinate conditions
that recast the data into a CTS form.
The lapse function is fixed by Eq.~(\ref{isco2}), which is derived from the condition for
$\Sigma_t$ to be a maximal slice. The shift vector is fixed, on the other hand,
by Eq.~(\ref{eq:recon}), which is needed to impose the 
quasi-equilibrium condition $\partial_t g_{ij}=0$.

\subsection{Komar Mass and Generalized Smarr Formula}

We need now to address how to fix the orbital 
angular velocity $\Omega$ in (\ref{eq:beta_infty}).
In GGB, $\Omega$ fixes the boundary condition that determines the 
initial data. Our case is different;
the initial data $(\phi,\,W^i)$ 
are obtained following the CTT method, so 
it is not until one solves for the shift vector that 
the issue of specifying a frame of reference emerges. 
If one chooses to solve for the shift with the asymptotic condition
$\beta^i = 2\,W^i$,
the data $(\phi,\,W^i,\,\alpha,\,\beta^i)$ 
should be viewed as computed in an inertial frame of reference.
 
One can transform these data to a rotating frame \cite{pedro} by simply changing the
shift vector to
\begin{equation}
\label{addrot}
\beta^i \rightarrow \beta^i + \Omega\,\left(\frac{\partial}{\partial\varphi_o}\right)^i\,.
\end{equation}
This transformation assumes that the inertial and rotating frame 
are instantaneously aligned. It also implies
that the lapse function, spatial metric and
extrinsic curvature are unchanged.

But the question still remains about how to fix $\Omega$.
One would like in principle that 
\begin{equation}
\label{kill_vec}
l^\mu = \left(\frac{\partial}{\partial t}\right)^\mu = \alpha\,N^\mu + \beta^\mu
\end{equation}
is a Killing vector, with $t$ the time coordinate associated with the rotating
frame. If such a Killing vector exists, it implies that the physical metric
and extrinsic curvature satisfy
\begin{eqnarray}
\partial_t \, \tilde g_{ij} &=& 0\\
\partial_t \, \tilde K_{ij} &=& 0\,.
\end{eqnarray}
As pointed out by GGB \cite{ggb1}, this requirement
is too strong since energy would have to be pumped continuously 
into the system to keep the binary at a fixed orbit, thus
violating asymptotic flatness. 

Instead, one can require that in the rotating frame the conformal metric and
trace of the extrinsic curvature satisfy
\begin{eqnarray}
\label{eq:cond1}
\partial_t \, g_{ij} &=& 0\\
\label{eq:cond2}
\partial_t \, K &=& 0\,.
\end{eqnarray}
Eq.~(\ref{eq:cond1}) is equivalent to the thin-sandwich
condition $u_{ij} = 0$.
Similarly, Eq.~(\ref{eq:cond2}) is nothing other than the condition
for $\Sigma_t$ to be a maximal slice. 

Given (\ref{eq:cond1}) and (\ref{eq:cond2}), as well as $K = 0$, the
evolution equations (\ref{etadot}) and (\ref{trkdot}) yield 
\begin{eqnarray}
\label{eq1}
(\tilde\Long\,\beta)_{ij}- 2\,\alpha\,\tilde A_{ij} &=&0\\
\label{eq2}
\widetilde\nabla^i\widetilde\nabla_i\,\alpha
- \alpha\,\tilde A^{ij}\,\tilde A_{ij} &=& 0\,,
\end{eqnarray} respectively.
Substitution of (\ref{eq1}) into (\ref{eq2}), together with
the momentum constraint $\widetilde\nabla_j\tilde A^{ij} = 0$, yields
\begin{equation}
\label{smarr1}
\widetilde\nabla^i(\widetilde\nabla_i\alpha - \tilde A_{ij}\,\beta^j) = 0\,.
\end{equation}
It is important to emphasize that 
this equation does not require conformal flatness to hold. 

Integration of (\ref{smarr1}) over $\Sigma_t$ excluding the black hole
singularities (sources) yields \cite{ggb1}
\begin{equation}
\label{smarr2}
\frac{1}{4\,\pi}\oint_{\infty}(\widetilde\nabla_i\alpha - \tilde A_{ij}\,\beta^j)\,dS^i
+ I = 0
\end{equation}
where
\begin{equation}
\label{I_int}
I \equiv \frac{1}{4\,\pi}
\sum_{A=1,2}\, \oint_{{\cal S}_A}(\widetilde\nabla_i\alpha 
- \tilde A_{ij}\,\beta^j)\,dS^i \,.
\end{equation}
Here ${\cal S}_A$ are 2-spheres surrounding the black holes with
$dS^i$ oriented toward the interior of the black hole.

The integral in (\ref{smarr2}) involving the gradient of the 
lapse has a physical interpretation. 
Given a time-like Killing vector $\xi^\mu$, 
the total mass (Komar mass) in a space-like hypersurface
is given by \cite{Komar}
\begin{equation}
\label{m_komar}
M_{KMR} =\frac{1}{4\,\pi} \oint_{\cal S} \widetilde
\nabla^\mu\xi^\nu\,n_\mu\,N_\nu\,dS\,,
\end{equation}
with $n^\mu$ a unit vector in $\Sigma_t$ normal to a 2-surface ${\cal S}$
enclosing  the sources. $N^\mu$ is the time-like unit normal to $\Sigma_t$.
In particular, one could take
${\cal S}$ to be at spatial infinity  
and $\xi^\mu \rightarrow \alpha\,(\partial/\partial t_o)^\mu$,
with $(\partial/\partial t_o)^\mu$ the time-translation
Killing vector of the flat metric. One can then show 
\cite{gb94} that 
\begin{equation}
\label{m_komar2}
M_{KMR} =\frac{1}{4\,\pi} \oint_{\infty}\widetilde\nabla_i\alpha\,dS^i\,.
\end{equation}
Thus, Eq.~(\ref{smarr2}) becomes
\begin{equation}
\label{smarr3}
M_{KMR} -
\frac{1}{4\,\pi}\oint_{\infty}\tilde A_{ij}\,\beta^j\,dS^i
+ I = 0\,.
\end{equation}
One must emphasize that here $M_{KMR}$ is meant to represent 
the result of the integral (\ref{m_komar2}). It does 
not imply the existence of a global time-like Killing vector in $\Sigma_t$.
That is, in general $\partial_t \phi \ne 0$
and $\partial_t \bar A_{ij} \ne 0$.

The second term in (\ref{smarr3}) has also a physical interpretation.
If $\beta^i \rightarrow \Omega\,(\partial/\partial\varphi_o)^i$, 
\begin{equation}
\label{jadm}
2\,\Omega\,J_{ADM}=\frac{1}{4\,\pi}\oint_{\infty}\tilde A_{ij}\,\beta^j\,dS^i\,,
\end{equation}
with $J_{ADM}$ the total angular momentum in $\Sigma_t$ \cite{york79}. 
Therefore,
\begin{equation}
\label{smarr4}
M_{KMR} - 2\,\Omega\,J_{ADM}+I = 0\,.
\end{equation}
One can then use this equation to solve for $\Omega$.
However, Eq.~(\ref{smarr4}) does not fix $\Omega$ in the sense of providing a
quasi-stationary situation. Given any solution 
$(\phi,\, \alpha,\, \beta^i)$ to the 
system of equations (\ref{isco1}--\ref{iscoshift}),
one can always solve for $\Omega$ from (\ref{smarr4}).
The value of $\Omega$ obtained in this fashion yields 
coordinate conditions$(\alpha,\,\beta^i)$ such that
Eqs.~(\ref{eq:cond1}) and (\ref{eq:cond2}) are satisfied.
It does not, however, necessarily imply that the vector $l^\mu$ 
in (\ref{kill_vec}) is an approximate helical Killing vector.

As pointed out by Cook \cite{cook2}, perhaps the most valuable
aspect of the work by GGB is to fix 
$\Omega$ by requiring 
that the ADM and Komar masses are equal. 
With this condition, Eq.~(\ref{smarr4}) becomes
\begin{equation}
\label{smarr5}
M_{ADM} - 2\,\Omega\,J_{ADM} + I = 0\,,
\end{equation}
with
\begin{equation}
\label{m_adm}
M_{ADM} = -\frac{1}{2\,\pi} \oint_{\infty}\widetilde\nabla_i\phi\,dS^i\,.
\end{equation} 
Eq.~(\ref{smarr5}) is also known as the generalized Smarr formula
\cite{ggb1}.

The idea behind constructing data with
$M_{ADM} = M_{KMR}$ is motivated by Beig's observation \cite{beig} that
for stationary space-times these masses are indeed equal.
Therefore, specifying $\Omega$ to fulfill $M_{ADM} = M_{KMR}$ would 
hopefully yield 
a co-rotating frame for quasi-circular orbits that 
minimizes a suitable norm of
$\partial_t \phi$ and $\partial_t \bar A_{ij}$ \cite{BLT}. 

If the shift vector solution to Eq.~(\ref{iscoshift}) is such that a
residual transverse-traceless part $u_{ij} = \partial_t \, g_{ij}$ is present, 
it is still possible to derive an expression similar to
(\ref{smarr5}). In this case, one needs to modify $I$ to be \cite{BLT}
\begin{eqnarray}
I &\equiv& \frac{1}{4\,\pi}
\sum_{A=1,2}\, \oint_{{\cal S}_A}(\widetilde\nabla_i\alpha 
- \tilde A_{ij}\,\beta^j)\,dS^i \\
&+&  \frac{1}{8\,\pi} \int \tilde A^{ij}\,\tilde g^{1/3}\,u_{ij} \sqrt{\tilde g}\, d^3x\,.
\end{eqnarray}
Even for non-vanishing $u_{ij}$, 
the condition $M_{ADM} = M_{KMR}$ can be applied. It will
minimize the norm of $\partial_t\,g_{ij}$, in addition to those
of $\partial_t \phi$ and $\partial_t \bar A_{ij}$ \cite{BLT}.
In Ref.~\cite{tichy}, this methodology was used to construct 
binary black hole sequences with punctures. The results are in close 
agreement with those by Baumgarte \cite{baumgarte}.
 
It is important to mention that in the CTS initial data construction
by GGB, Eq.~(\ref{smarr5}) is only used as a 
consistency check. The GGB methodology
is such that part of the outcome is the rotating frame
itself.
As discussed before, 
the angular velocity $\Omega$ of the co-rotating frame
enters in the boundary conditions needed to solve for the
shift vector. That is, $\Omega$ is adjusted to yield $M_{ADM} = M_{KMR}$.
In our approach, on the other hand, the initial data is given
by the CTT method, with $\Omega$ a derived quantity. To fix $M_{ADM} = M_{KMR}$,
one adjusts instead \cite{BLT} the parameters $a$ and $b$ in 
(\ref{phi_asymp}) and (\ref{alpha_asymp}).

Finally, if instead of  
$\beta^i \rightarrow \Omega\,(\partial/\partial\varphi_o)^i$, 
one has that 
$\beta^i \rightarrow V\,(\partial/\partial x_o)^i$, 
with $(\partial/\partial x_o)^i$ the Killing vector of the flat
metric along the direction $x^i_o$, the integral in the
the second term in (\ref{smarr3}) has a different physical interpretation.
In this case,
\begin{equation}
\label{padm}
\frac{1}{4\,\pi}\oint_{\infty}\tilde A_{ij}\,\beta^j\,dS^i = 2\,V\,P_{ADM}\,,
\end{equation}
with $P_{ADM}$ the total linear momentum in $\Sigma_t$ \cite{york79}. 
The Smarr formula reads in this situation
\begin{equation}
\label{smarr6}
M_{KMR} - 2\,V\,P_{ADM}+I = 0\,.
\end{equation}

\section{CTS Punctures}
\label{sec:punctures}

We will now explicitly consider CTS data involving punctures \cite{puncture}.
One decomposes $\phi$ and $\alpha\,\phi$ as
\begin{eqnarray}
\label{puncture1}
\phi &=& u + \frac{1}{p}\\
\label{puncture2}
\alpha\,\phi &=& v + \frac{1}{q}\,,
\end{eqnarray}
such that
\begin{eqnarray}
\frac{1}{p} &=&  \frac{a_1}{r_1} +\frac{a_2}{r_2}\\ 
\frac{1}{q} &=& -\frac{b_1}{r_1} -\frac{b_2}{r_2}\,, 
\end{eqnarray}
respectively. Above, $a$ and $b$  
are parameters and $u$ and $v$ regular functions.
The parameter $a$ and $b$ can be related to the ``bare"
masses of the punctures \cite{BLT}. 

Substitution of (\ref{puncture1}) and (\ref{puncture2}) into
Eqs.~(\ref{cts:hc}) and (\ref{eq:lapse2}) yields
\begin{eqnarray}
\label{eq:uno}
\nabla^2 u &=&  -\frac{\phi^5}{32\,\alpha^2} (\Long \beta)^2 \\
\label{eq:dos}
\nabla^2 v &=&   \frac{7}{32}\,\frac{\phi^6}{(\alpha\,\phi)}\, (\Long \beta)^2\,.
\end{eqnarray}
For asymptotic flatness, the following Robin boundary conditions
are imposed far from the holes:
\begin{eqnarray}
n^i\,\nabla_i u &=& \frac{(1-u)}{r}\\
n^i\,\nabla_i v &=& \frac{(1-v)}{r}\,.
\end{eqnarray}

As pointed out in Ref.~\cite{hannam}, the problem with
punctures in the CTS method is that a natural choice is 
\begin{eqnarray}
\lim_{r\rightarrow\infty} \alpha &=& 1\\
\lim_{r\rightarrow r_A} \alpha &=& -c_A < 0\,.
\end{eqnarray} 
Therefore, the lapse function would necessarily have to vanish at some
internal boundary. In order to have regularity of $u$ and $v$, one would 
have to have that $(\Long\beta)^{ij}$ vanishes at least as fast as $\alpha$ at
that internal boundary. Imposing the vanishing of $(\Long\beta)^{ij}$ in general
requires more freedom than is available. One could in principle 
set $c_A < 0$. This choice, however, yields solutions far from being
quasi-stationary.

If the combination of punctures and CTS seems to be intrinsically 
troublesome, it seems then puzzling that 
CTS conformal-imaging and CTT puncture binary black hole initial data sets 
exhibit good agreement as the separation of the binary increases.
It should then be
possible to translate, using the procedure described in the previous section,
the CTT-puncture data into a CTS form in spite of the presence of an
internal boundary where $\alpha=0$. 
To demonstrate that this is the case, we consider single black hole solutions.
For large separations, not only does the solution approximate that of a single,
spinning black hole far away from the binary system, but it also approaches
the solution of a single black hole with linear momentum near each of the
black holes. Specifically, we will consider Bowen and York \cite{BY}
single black hole solutions using punctures and show that it is possible to
bring those solutions into a CTS form.

From (\ref{eq:uno}), (\ref{eq:dos}) and (\ref{eq:recon}), the equations to be solved are:
\begin{eqnarray}
\nabla^2 u &=&  -\frac{1}{8}\frac{1}{(p\,u+1)^7}\,p^7(\Long W)^2 \\
\nabla^2 v &=&   \frac{7}{8}\frac{(q\,v+1)}{(p\,u+1)^8}\,\frac{p^8}{q}\, (\Long W)^2\\
(\Long \beta)^{ij} &=& 2\,\frac{(q\,v+1)}{(p\,u+1)^7}\,\frac{p^7}{q}\,(\Long W)^{ij}\,.
\end{eqnarray}
with $(\Long W)^{ij}$ given by (\ref{alin}) or (\ref{aspin}).
Notice that because $(\Long W)^{ij}$ is linear in the momentum, 
we will only consider perturbative solutions \cite{pullin1,pullin2}
to second order in the 
momentum for $u$ and $v$ and linear order for $\beta^i$. 
Therefore,
\begin{eqnarray}
\label{cts_by1}
\nabla^2 u &=&  -\frac{1}{8}\frac{\bar r^7}{(\bar r+1)^7}\,(\Long W)^2 \\
\label{cts_by2}
\nabla^2 v &=&   \frac{7}{8}\frac{(\bar r-\bar b)\,\bar r^7}{(\bar r+1)^8}\,(\Long W)^2\\
\label{cts_by3}
(\Long \beta)^{ij} &=& 2\,\frac{(\bar r-\bar b)\,\bar r^6}{(\bar r+1)^7}\,(\Long W)^{ij}\,,
\end{eqnarray}
where over-bars here denote scaling with respect to the parameter 
$a$ (i.e. $\bar r \equiv r/a$, $\bar b = b/a$, $\bar P = P/a$ and $\bar J = J/a^2$).

\subsection{Spinning Bowen-York Black Hole}
\label{sec:spinning_punct}

For a spinning Bowen-York black hole \cite{BY},
the only non-vanishing component of $W^i$ and $(\Long W)_{ij}$ are
\begin{eqnarray}
W^\varphi &=& -\frac{J}{r^3}\\
(\Long W)_{r\varphi} &=& \frac{3\,J}{r^2}\sin^2\theta \,,
\end{eqnarray}
where $J$ is the total angular momentum. Therefore,
\begin{equation}
\label{lw_spin}
(\Long W)^2 = \frac{18\,J^2}{r^6}\sin^2\theta =\frac{12\,J^2}{r^6}\,[P_0(\mu)-P_2(\mu)] \,,
\end{equation}
where $\mu \equiv \cos\theta$ and $P_l$ are Legendre polynomials.
Substitution of (\ref{lw_spin})
into Eqs.~(\ref{cts_by1}--\ref{cts_by3}) yields
\begin{eqnarray}
\label{cts_spin1}
\nabla^2 u &=&  -\frac{3}{2}\frac{J^2\,r}{(r+1)^7}\,[P_0(\mu)-P_2(\mu)] \\
\label{cts_spin2}
\nabla^2 v &=&   \frac{21}{2}\frac{J^2\,r\,(r-b)}{(r+1)^8}\,[P_0(\mu)-P_2(\mu)]\\
\label{cts_spin3}
\partial_r\beta^\varphi &=& 6\,J\,r^2\,\frac{(r-b)}{(r+1)^7}\,.
\end{eqnarray}
To simplify notation,
we have dropped the over-bars, and rescaled 
$\beta^\varphi$ as $\beta^\varphi\,a$.

Regular solutions to (\ref{cts_spin1})-(\ref{cts_spin3}) are
\begin{eqnarray}
u(r,\,\mu) &=& J^2\,[\,u_0(r)\,P_0(\mu)+u_2(r)\,r^2\,P_2(\mu)\,]\\
v(r,\,\mu) &=& J^2\,[\,v_0(r)\,P_0(\mu)+v_2(r)\,r^2\,P_2(\mu)\,]\\
\beta^\varphi(r) &=& J\,\Bigg[\frac{-2}{(r+1)^3}
+\frac{3\,(b+3)}{2\,(r+1)^4}\nonumber\\
&-& \frac{6\,(2\,b+3)}{5\,(r+1)^5}
+\frac{(b+1)}{(r+1)^6}+\frac{(1-b)}{10}\Bigg]\,,
\end{eqnarray}
where
\begin{eqnarray}
40\,u_0(r) &=& \frac{1}{(r+1)} +  \frac{1}{(r+1)^2} +  \frac{1}{(r+1)^3} \nonumber\\
           &-& \frac{4}{(r+1)^4} + \frac{2}{(r+1)^5} \\
20\,u_2(r) &=& \frac{-1}{(r+1)^5}\\
40\,v_0(r) &=& \frac{(3\,b-4)}{(r+1)} +  \frac{(3\,b-4)}{(r+1)^2}\nonumber\\
           &+&\frac{(3\,b-4)}{(r+1)^3}+ \frac{(3\,b+31)}{(r+1)^4}\nonumber\\
           &-& \frac{(18\,b+32)}{(r+1)^5}+ \frac{10\,(b+1)}{(r+1)^6} \\
20\,v_2(r) &=& \frac{r\,(6-b)-6\,b+1}{(r+1)^6}\,.
\end{eqnarray} 

These solutions have the following asymptotic form
when $r\rightarrow\infty$:
\begin{eqnarray}
\phi &=& 1+\frac{1}{r}\,\left(1+\frac{J^2}{40}\right)\\
\alpha &=& 1-\frac{1}{r}\,\left[b+1 + \frac{J^2\,(5-3\,b)}{40}\right]\\
\beta^\varphi &=&J\,\left[\frac{(1-b)}{10}-\frac{2}{r^3}\right]\nonumber\\
              &=&\frac{J\,(1-b)}{10}+2\,W^\varphi\,.
\end{eqnarray}
Furthermore, integration constants for $\beta^\varphi$ have been chosen
such that $\beta^\varphi(0) = 0$. 
From Eqs.~(\ref{m_adm}) and (\ref{m_komar2}), 
the ADM and Komar masses are respectively
\begin{eqnarray}
M_{ADM} &=& 2 + \frac{J^2}{20}\\
M_{KMR} &=& b+1 + \frac{J^2\,(5-3\,b)}{40}
\end{eqnarray}

On the other hand, it is not difficult to show that near the
puncture the integral $I$, defined by (\ref{I_int}), yields
\begin{eqnarray}
I &=& -\left[ 1+b + J^2\,(b\,u_0(0)+v_0(0))\right]\nonumber\\
  &=& -\left[ 1+b + \frac{J^2\,(5\,b-3)}{40}\right] \,.
\end{eqnarray} 
Similarly from (\ref{jadm}),
\begin{equation}
\Omega =\frac{J\,(1-b)}{10}\,,
\end{equation}
which is consistent with our requirement
that asymptotically $\beta^\varphi\rightarrow\Omega$. 
Clearly, the computed values for $M_{KMR}$, $I$ and and $\Omega$
satisfy the Smarr formula $M_{KMR} - 2\,\Omega\,J + I=0$.
In addition, to satisfy the condition $M_{ADM} = M_{KMR}$,
one needs to set $b=1$, which in turn implies that $\Omega=0$.

\subsection{Boosted Bowen-York Black Hole}
\label{sec:boosted_punct}

The only non-vanishing components of $W^i$ and $(\Long W)_{ij}$ for
a boosted Bowen-York black hole \cite{BY} are:
\begin{eqnarray}
W^r &=& -2\,\frac{P}{r}\,\cos\theta\nonumber\\
W^\theta &=& \frac{7}{4}\frac{P}{r^2}\,\sin\theta\\
(\Long W)_{rr}             &=&            3\,\frac{P}{r^2}\,\cos\theta \nonumber\\
(\Long W)_{r\theta}        &=& -\frac{3}{2}\,\frac{P}{r}\,\cos\theta \nonumber\\
(\Long W)_{\theta\theta}   &=& -\frac{3}{2}\,P\,\cos\theta \nonumber\\
(\Long W)_{\varphi\varphi} &=& -\frac{3}{2}\,P\,\cos\theta \,\sin^2\theta\,,
\end{eqnarray}
where the linear momentum is assumed to point along the positive $z$ axis
and have magnitude $P$.

With the above expression, the term $(\Long W)^2$ takes the form
\begin{equation}
(\Long W)^2 = \frac{3}{2}\,\frac{P}{r^4}\,[\,5\,P_0(\mu)+4\,P_2(\mu)\,]\,.
\end{equation}
The equations to be solved for $u$ and $v$ in this case are:
\begin{eqnarray}
\label{cts_boost1}
\nabla^2 u &=&  -\frac{3}{16}\frac{P^2\,r^3}{(r+1)^7}\,[5\,P_0(\mu)+4\,P_2(\mu)] \\
\label{cts_boost2}
\nabla^2 v &=&   \frac{21}{16}\frac{P^2\,r^3\,(r-b)}{(r+1)^8}\,[5\,P_0(\mu)+4\,P_2(\mu)]\,,
\end{eqnarray}
and for the shift vector
\begin{eqnarray}
\label{cts_boost3}
r\,\partial_r\left(\frac{\beta^r}{r}\right) - \partial_\theta\beta^\theta 
 &=& 9\,\frac{P\,r^4\,(r-b)}{(r+1)^7}\,\cos\theta\\
\label{cts_boost4}
r^2\,\partial_r\beta^\theta + \partial_\theta\,\beta^r 
 &=&-3\,\frac{P\,r^5\,(r-b)}{(r+1)^7}\,\sin\theta\,.
\end{eqnarray}
In the equations above, we have introduced the same scaling with respect to
the parameter $a$ as in the previous section.

Regular solutions to (\ref{cts_boost1})--(\ref{cts_boost4}) are:
\begin{eqnarray}
u(r,\,\mu) &=& P^2\,[\,u_0(r)\,P_0(\mu)+u_2(r)\,P_2(\mu)\,]\\
v(r,\,\mu) &=& P^2\,[\,v_0(r)\,P_0(\mu)+v_2(r)\,P_2(\mu)\,]\\
\beta^r(r,\,\theta) &=& P\,R(r)\,\cos\theta\\
\beta^\theta(r,\,\theta) &=& P\,S(r)\,\frac{1}{r}\,\sin\theta\,,
\end{eqnarray}
where
\begin{eqnarray}
\frac{32}{5}\,u_0(r) &=& \frac{1}{(r+1)} -  \frac{2}{(r+1)^2} +  \frac{2}{(r+1)^3} \nonumber\\
           &-& \frac{1}{(r+1)^4} + \frac{1}{5\,(r+1)^5} \\
80\,r\,u_2(r) &=& \frac{15}{(r+1)} +  \frac{132}{(r+1)^2}\nonumber\\
           &+&\frac{53}{(r+1)^3}+ \frac{96}{(r+1)^4}\nonumber\\
           &+& \frac{82}{(r+1)^5}+ \frac{84}{r\,(r+1)^5}\nonumber \\
           &-& \frac{84}{r^2}\,\ln(r+1)\\
\frac{32}{5}\,v_0(r) &=& \frac{(b-6)}{(r+1)} +  \frac{(b+15)}{(r+1)^2}\nonumber\\
           &-&\frac{(6\,b+20)}{(r+1)^3}+ \frac{(8\,b+15)}{(r+1)^4}\nonumber\\
           &-& \frac{(23\,b+30)}{5\,(r+1)^5}+ \frac{(b+1)}{(r+1)^6} \\
80\,r^2\,v_2(r) &=& -\frac{(201\,b+1020)}{(r+1)} +  \frac{(275\,b+825)}{(r+1)^2}\nonumber\\
           &-&\frac{(287\,b+588)}{(r+1)^3}+ \frac{(185\,b+283)}{(r+1)^4}\nonumber\\
           &-& \frac{(66\,b+80)}{(r+1)^5}+ \frac{10\,(b+1)}{(r+1)^6}\nonumber \\
           &+& \frac{84}{r}\,(b+8)\,\ln(r+1)-105\\
R(r) &=& \frac{1}{(r+1)^6}\,\Big[
             \frac{1}{8}        (b-5)
            +\frac{3}{4}        (b-5)\,r\nonumber\\
     &+&     \frac{1}{80}(151\,b-751)\,r^2\nonumber\\ 
     &+&     \frac{1}{40}(103\,b-503)\,r^3\nonumber\\ 
     &+&     \frac{3}{16}   (11\,b-51)\,r^4 
            - 4\,r^5\Big]\nonumber\\
     &-&\frac{1}{8}(b-5)\\
S(r) &=& \frac{1}{(r+1)^6}\,\Big[
            -\frac{1}{8}        (b-5)
            -\frac{3}{4}        (b-5)\,r\nonumber\\
     &-&     \frac{1}{80}(149\,b-749)\,r^2\nonumber\\ 
     &-&     \frac{1}{40}(97\,b-497)\,r^3\nonumber\\ 
     &-&     \frac{3}{16}   (9\,b-49)\,r^4 
            + \frac{7}{2}\,r^5\Big]\nonumber\\
     &+&\frac{1}{8}(b-5)\,.
\end{eqnarray}

As $r\rightarrow\infty$, these solutions have the following form:
\begin{eqnarray}
\phi &=& 1+\frac{1}{r}\,\left(1+\frac{5\,P^2}{32}\right)\\
\alpha &=& 1-\frac{1}{r}\,\left[b+1 + \frac{5\,P^2\,(7-b)}{32}\right]\\
\beta^r &=&-P\,\left[\frac{(b-5)}{8}+\frac{4}{r}\right]\,\cos\theta\nonumber\\
 &=&-P\,\frac{(b-5)}{8}\,\cos\theta + 2\,W^r\label{beta_r}\\
\beta^\theta &=& P\,\left[\frac{(b-5)}{8}+\frac{7}{2\,r}\right]\frac{\sin\theta}{r}\nonumber\\
 &=& P\,\frac{(b-5)}{8}\frac{\sin\theta}{r}+2\,W^\theta\label{beta_t}\,.
\end{eqnarray} 
From (\ref{beta_r}) and (\ref{beta_t}), one has that asymptotically
\begin{equation}
\beta^i = \frac{(5-b)}{8}\,P^i + 2\,W^i\,,
\end{equation}
where $W^i$ is given by (\ref{wlin}).

From Eqs.~(\ref{m_adm}) and (\ref{m_komar2}), 
the ADM and Komar masses are respectively
\begin{eqnarray}
M_{ADM} &=& 2+\frac{5\,P^2}{16}\\
M_{KMR} &=& b+1 + \frac{5\,P^2\,(7-b)}{32}\,,
\end{eqnarray}
and from (\ref{jadm}),
\begin{equation}
\label{boost_int}
V= \frac{(5-b)}{8}P\,.
\end{equation}
At the puncture, it is not difficult to show that 
\begin{eqnarray}
I &=& -\left[1+b + P^2\,(b\,u_0(0)+v_0(0))\right]\nonumber\\
   &=& -\left[1+b+\frac{P^2\,(3\,b-5)}{32}\right] \,.
\end{eqnarray}
It follows then that $M_{KMR}$, $I$ and $V$ above satisfy, as in the
case of a spinning puncture, 
the Smarr relationship $M_{KMR} - 2\,V\,P + I = 0$.
If in addition one
imposes the condition $M_{ADM}=M_{KMR}$, then $b=1-20\,P^2/32+O(P^4)$
and $V = P/2+O(P^3)$.   

\begin{figure}
\includegraphics[height=7cm,width=8cm,angle=-90]{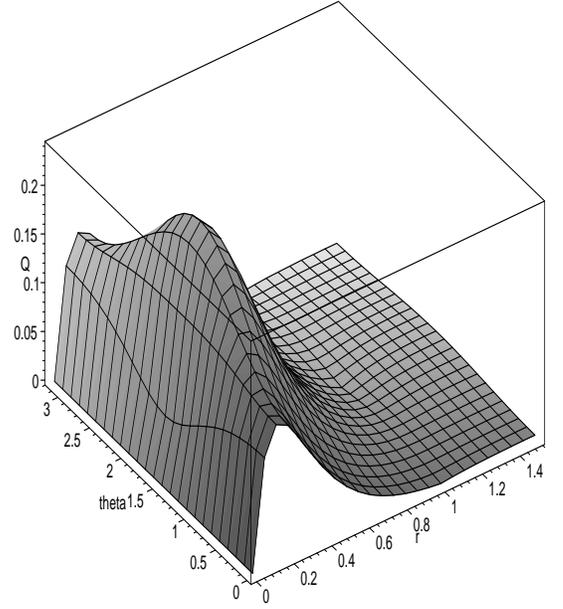}
\caption{
Surface plot of $Q \equiv 
(\partial_t\tilde A_{ij}\partial_t\tilde A_{kl}g^{ik}g^{jl})^{1/2}$
for a spinning puncture in units of $J^2$.} 
\label{fig1}
\end{figure}

\begin{figure}
\includegraphics[height=7cm,width=8cm,angle=-90]{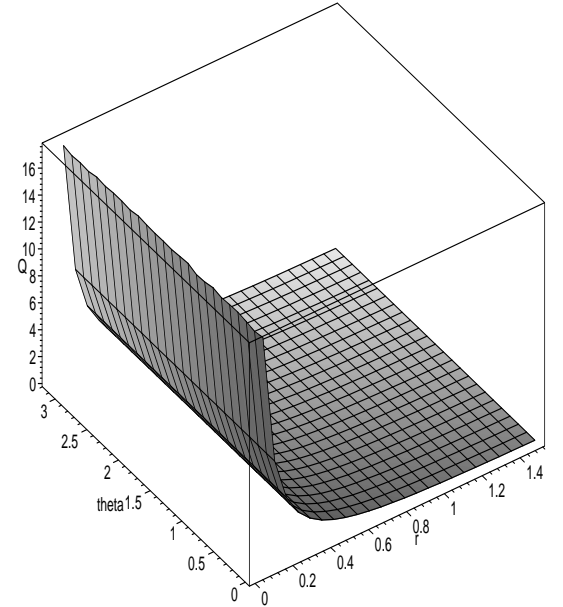}
\caption{Surface plot of $Q \equiv 
(\partial_t\tilde A_{ij}\partial_t\tilde A_{kl}g^{ik}g^{jl})^{1/2}$
for a boosted puncture in units of $P^2$.} 
\label{fig2}
\end{figure}

\subsection{Approximate Killing Vectors}

In the previous two sections, we showed that it is possible, at least
at the perturbative level, to bring single, spinning or boosted CTT-punctures
into a CTS form. By construction, the data satisfy the conditions
$\partial_t g_{ij} = 0$ and $\partial_t K = 0$. Furthermore,
in these cases the problems associated
with the vanishing of the lapse are avoided. What remains is to investigate
how close the data come to satisfying the conditions $\partial_t \phi = 0$ and
$\partial_t\tilde A_{ij} = 0$, conditions necessary for the existence of
an exact Killing vector.  

From Eqs.~(\ref{phidot}) and (\ref{adot}), we have that
\begin{eqnarray}
\partial_t \phi &=& \beta^i\,\nabla_i\,\phi
+ \frac{1}{6}\,\phi\,\nabla_i\beta^i \label{phidot2}\\ 
\partial_t\,\tilde A_{ij} &=& {\cal L}_\beta \tilde A_{ij}  
- 2\,\alpha\,\tilde A_{ik}\,\tilde A^k\,_j \nonumber \\
&+& [\alpha\,\tilde R_{ij}-\widetilde\nabla_i\widetilde\nabla_j\alpha]^{TF} 
\,, \label{adot2}
\end{eqnarray}
where the condition $K=0$ has been used.
Because the solutions in Secs.~\ref{sec:spinning_punct} and \ref{sec:boosted_punct}
are up to second order in the momentum (linear or angular) for $\phi$ and $\alpha$ and
first order for $\beta^i$, one only needs to focus attention to
$\partial_t \phi$ and $\partial_t\tilde A_{ij}$ up to second order
in the momentum.

From Sec.~\ref{sec:spinning_punct}, we have that for a spinning puncture
the solutions are independent of the coordinate $\varphi$ and 
the only non-vanishing component of the 
shift vector is $\beta^\varphi$. It is then clear from
(\ref{phidot2}) that $\partial_t \phi = 0$. For the case of a
boosted puncture, on the other hand, one has that
\begin{equation}
\partial_t\phi = \frac{P}{80}\frac{r\,\cos\theta}{(r+1)^5}(b-1)(10\,r^2+5\,r+1)+O(P^3)\,.
\end{equation}
At the end of Sec.~\ref{sec:boosted_punct}, we saw that in order to
have $M_{ADM}=M_{KMR}$, one needs $b=1+O(P^2)$; therefore,
$\partial_t\phi = O(P^3) \approx 0$. To seoncd order in the momentum, 
$\partial_t\phi$ vanishes identically.

To zero order in $P$ or $J$, Eq.~(\ref{adot2}) equation reads
\begin{equation}
\label{eq:xxx}
\partial_t\,\tilde A_{ij} = 
\frac{(b-1)}{r\,(r+1)^2}\hbox{diag}[-2,\,r^2,\,r^2\,\sin\theta^2]\,,
\end{equation} 
Therefore, $\partial_t\,\tilde A_{ij} = 0$ since to this order $M_{ADM}=M_{KMR}$ 
yields $b=1$. To first order in $P$ and $J$, Eq.~(\ref{adot2}) vanishes identically. 

On the other hand, to second order in $J$, we have that for a spinning puncture,
even for $b=1$, the only 
vanishing components are $\partial_t\tilde A_{r \varphi}$ and
$\partial_t\tilde A_{\theta \varphi}$.
Similarly, to second order in $P$, these same components of $\partial_t\tilde A_{ij}$ 
are the only ones vanishing,
even if $b=1-20\,P^2/32$ as required by $M_{KMR} = M_{ADM}$.
The non-vanishing of $\partial_t\tilde A_{ij}$ is more apparent from
Fig.~\ref{fig1} and {\ref{fig2} where we plot the quantity 
$Q \equiv (\partial_t\tilde A_{ij}\partial_t\tilde A_{kl}g^{ik}g^{jl})^{1/2}$
as a function of the coordinates $r$ and $\theta$
for both cases. 

Having $\partial_t\,\tilde A_{ij} \ne 0$ to second order on $P$ and $J$
implies that the Bowen-York puncture solutions do not have an exact Killing vector
to this order.
This is consistent with the results in Refs.~\cite{pullin1,pullin2} showing 
that Bowen-York initial data for a spinning or boosted black hole do not
represent a constant-time slice of a Kerr or a boosted Schwarzschild black 
hole, respectively. 

\section{Conclusions}
\label{sec:conclusions}

We have considered single boosted or spinning black hole punctures of the
Bowen-York type and demonstrated that, at the perturbative level, it is 
possible to establish a direct connection
between CTT and CTS initial data satisfying the conditions used by GGB
needed for the existence of an approximate Killing vector. 
Since for widely separated binary systems the black holes can be viewed as
nearly isolated, our results contribute to explaining the agreement between
CBB and GGB binary data sets as the separation of the binary system increases. 

\section{Acknowledgments}
This work was supported by NSF grants
PHY-9800973 and PHY-0114375.
Special thanks to Abhay Ashtekar, Bernd Br\"{u}gmann and Wolfgang Tichy 
for helpful discussions.  Thanks also to Bernard Kelly for carfully reviewing 
the manuscript.
Work supported in part by
the Center for Gravitational Wave Physics funded
by the National Science Foundation under Cooperative Agreement PHY-0114375.

\end{document}